\documentclass[preprint2]{aastex}

\usepackage{epsf}
\usepackage{graphics}
\usepackage{color}

\newcommand{\be}{\begin{equation}}
\newcommand{\ee}{\end{equation}}

\newcommand{\bx}{$\beta_{\rm X}$}

\newcommand{\plm}{$\pm$}
\newcommand{\nh}{$N_{\rm H}$}
\newcommand{\swift}{\mbox{\it Swift}}	    

\newcommand{\nob}{55}  
\shorttitle{Redshift Filtering by Swift XRT Column Density}
\shortauthors{Grupe et al.}

\begin{document}


\def\etal{{\it et\thinspace al.}\ }
\def\alp{{$\alpha$}\ }
\def\al2{{$\alpha^2$}\ }


\title{Redshift Filtering by {\it Swift} Apparent X-ray Column Density
}


\author{Dirk Grupe\altaffilmark{1},\email{grupe@astro.psu.edu} 
John A. Nousek\altaffilmark{1},
Daniel E. vanden Berk\altaffilmark{1},
Peter W.A. Roming\altaffilmark{1}, 
David N. Burrows\altaffilmark{1}, 
Olivier Godet\altaffilmark{2},
Julian Osborne\altaffilmark{2},
Neil Gehrels\altaffilmark{3}
}

\altaffiltext{1}{ Astronomy Department, Pennsylvania State
University, 525 Davey Lab, University Park, PA 16802} 
\altaffiltext{2}{Department of Physics \& Astronomy, University of Leicester,
Leicester, LE1 7RH, UK}
\altaffiltext{3}{Astrophysics Science Division, Astroparticle Physics Laboratory,
Code 661, NASA Goddard Space Flight Center, Greenbelt, MD 20771 }




\begin{abstract}
We remark on the utility of an observational relation between the 
absorption column density in excess of the Galactic absorption column density,
$\Delta N_{\rm H} = N_{\rm H, fit} - N_{\rm H, gal}$,
and redshift, z, determined from all \nob\  \swift-observed long bursts with 
spectroscopic redshifts as of 2006 December. 
The absorption column densities, $N_{\rm H, fit}$,
are determined from powerlaw fits
to the X-ray spectra with the absorption column density left as a free
parameter. 
We find that higher excess
absorption column densities with $\Delta N_{\rm H} > 2\times 10^{21}$ cm$^{-2}$
are only present in bursts
with redshifts z$<$2. Low absorption column densities with 
$\Delta N_{\rm H} < 1\times 10^{21}$ cm$^{-2}$
appear preferentially in
high-redshift bursts. 
Our interpretation is that
this relation between redshift and excess column density
is an observational effect resulting from the shift of the source rest-frame
energy range below 1 keV out of the XRT observable energy range for high
redshift bursts. 
We found a clear anti-correlation between $\Delta N_{\rm H}$ and z that can be used
to estimate the range of the maximum redshift of an afterglow. 
A critical application of our finding is that rapid X-ray
observations can be used to optimize the instrumentation used for 
ground-based optical/NIR follow-up observations. Ground-based spectroscopic
redshift measurements of as many bursts as possible are crucial for GRB science.  
\end{abstract}

\keywords{GRBs: general
}

\section{Introduction}
The \swift\ mission \citep{gehrels04} has revolutionized the study of
Gamma-Ray Burst (GRB) afterglows.  The mission, which
relies upon training sensitive X-ray and optical telescopes on new
GRBs as rapidly as possible, has resulted in the accurate positioning
of GRB afterglows on a timescale of minutes.  Especially for long GRBs,
this rapid localization has proven highly effective at identifying afterglows
for study at other wavelengths.  Already, in about 2 years of operation,
\swift\ has localized more than twice as many GRB afterglows than had
been localized in the eight years preceding \swift\ \citep{burrows06b}.

However, \swift\ alone cannot address all the important issues in GRB
research.  One of the most important GRB parameters is the redshift of a burst. 
Knowledge of the redshift is not only crucial for determining the 
luminosity and other physical parameters of the burst, but also permits
optimization of ground-based observations.
For example a determination of the redshift for high-redshift GRBs
requires large telescopes with infra-red sensitivity, because the Lyman absorption
edge gets shifted beyond the long wavelength end of the \swift-UVOT
sensitivity for redshifts above 5 \citep{roming06}.
GRBs fade rapidly and only spectroscopy can provide reliable redshifts.
Therefore,
to search for redshifts as high as GRB050904 \citep[z=6.29; ][]{kawai05},
or even to the unexplored $z\sim 7-10$ range, requires 
telescopes in the 8-10 m
class making observations within the first night or two of the Swift
discovery.
As listed in this paper,
about 6\% of all \swift-observed GRBs with spectroscopic redshift
have redshift z$>$5.

Observing time on such large, world-class telescopes is an extremely
precious commodity.  Currently the limited number of target of opportunity 
programs must triggered based
on very limited information in order to spectroscopically
observe the many discovered \swift\ GRB afterglow.  
 While nearly all promptly observed
\swift\ long GRBs can be localized by the X-ray Telescope
\citep[XRT; ][]{burrows05},
only about 34\% are detected by the
UV/Optical Telescope \citep[UVOT; ][]{roming05, roming06}.
Thus observers looking for high redshift bursts can filter the approximately
100 GRBs found by \swift\ each year, by using the criterion:  an
afterglow is detected by the XRT, but not detected by the UVOT.
Unfortunately this criterion alone only reduces the rate
of high-z candidates by about 1/3, leaving about 67 GRBs to be
observed per year.  Moreover, many other factors, e.g. reddening, can result in
suppression of afterglow emission in the UVOT sensitivity range
(see Roming et al. 2006 for a discussion).

Of course large telescope observers can wait to see if any of the
smaller robotic telescopes can select candidates based on the
broad-band photometric studies conducted by these telescopes.
Often the cost of doing this is to lose several hours waiting for
the results of these small telescope studies to be reduced and
transmitted, and to be subject to the vagaries of weather and other
observing constraints on the ground. Also GRB afterglows decay rapidly. Every
hour of waiting reduces the chances of obtaining an optical and/or NIR
spectrum of the afterglow and therefore significantly
decreases the chances of measuring the
redshift of the burst from a spectrum. 

To see whether Swift XRT data alone can help to provide very
early information, we conducted a study of all \nob\  Swift
long GRB afterglows with known redshifts by November 2006.  
On a timescale of
one or two hours after the initial XRT detection of a
new GRB, the XRT telemeters information
to the ground from which many properties of the burst can be determined,
including accurate positions, X-ray flux and X-ray spectral information.

The Swift team has been routinely analyzing these data, and
reporting them to the world via the GRB Coordinates Network 
\citep[GCN; ][]{barthelmy95}.
It has been found that the typical afterglow can be fit with a simple
power law model spectrum plus the effects of a variable amount
of absorbing material \citep[e.g. ][]{stratta04, campana06}.  This absorbing
material can be located either in our Galaxy, in the host galaxy
of the GRB, or in intervening gas clouds. As shown on a theoretical basis by
\citet{ramirez02}, extinction in GRBs is expected.  
Recently \citet{prochaska06} presented 
the results of high-resolution optical spectroscopy of the interstellar medium 
of GRB host galaxies of \swift\ GRB afterglows.
For low redshift bursts, in the observer's rest-frame,
the effects of local, and intervening absorption appear at roughly similar
energies, and thus the effects are intermixed. For high redshift bursts, the
absorbing material in the host galaxy will incur a substantial redshift.
The result is that the energy band of the Swift XRT is shifted to higher
 energies in the GRB rest frame, making it difficult to detect X-ray 
 absorption.  Without prior knowledge of the redshift, X-ray observations 
 result in measurements of NH that are systematically low compared to the 
 actual absorbing column in the host galaxy.

Thus if we take the apparent absorption column density $N_{\rm H, fit}$, using the 
photo-electric cross-section as given by \citet{morrison83}, 
in
the observer's rest-frame, and subtract off the known absorption column
density in our Galaxy \citep[$N_{\rm H, gal}$ as given by ][]{dic90}, the
residual $\Delta N_{\rm H} = N_{\rm H, intr} - N_{\rm H, gal}$
will reflect the redshifted column density either in
the source or in the intervening line-of-sight.  If this residual
column density appears high, then it is very likely that the
GRB is close, because distant GRB absorption effects are
overwhelmed by the redshift effect.  Of course if the
column density is low, we cannot tell whether the GRB is
near - with low intrinsic absorption - or far - with either
low intrinsic absorption or redshifted high absorption.
A similar method has also been proposed to estimate the redshifts of
high-redshift quasars \citep{wang04}.

We present the values for absorption measured in the
first orbit of data
for all \nob\  Swift GRBs with known spectroscopic redshifts (section 2).
The X-ray data are typically available 1-2 hours after the detection of the burst,
except for those few bursts for which observing constraints prevent \swift\ from
slewing immediately.
The relation we are proposing is not a functional
prediction (i.e. we do not suggest that GRB redshift is
derivable from XRT absorption), but instead we argue that
we can predict the maximum redshift of an afterglow on the basis of the excess
absorption $\Delta N_{\rm H}$.
   In section 3 we present the results. 
Finally, in section 4 we present our conclusion. 

Throughout
the paper spectral  index \bx\  
is defined as $F_{\nu}(\nu)\propto\nu^{-\beta_{\rm X}}$.  
All errors are 1$\sigma$ unless stated otherwise. 
 
\section{\label{observe} Observations and data reduction}

Table\,\ref{grb_list} lists all \nob\  long 
GRBs with reported spectroscopic redshifts (up to 2006 November)
which were observed
by \swift. We did not include short bursts 
since they  most likely have different physical processes
than the long bursts. Furthermore,
short burst have typically been detected at relatively low
redshifts \citep[e.g. ][]{berger06d}, even though \citet{levan06} suggested that the 
short GRB 060121 is possibly at z$>$4.5.
 We limit our sample to those bursts with spectroscopic
redshifts only, because these are the most reliable redshift measurements. Other
methods such as photometric redshifts are less certain because a drop-out in bluer
filters can also be caused by strong dust reddening.

The XRT data were reduced by the {\it xrtpipeline} software version 0.10.4 which
is part of the HEASOFT version 6.1.1.
For XRT Photon Counting mode data \citep[PC; ][]{hillj04},
source photons were selected by {\it XSELECT} version 2.4
in a circular region with a radius
of r=47$^{''}$ and the background photons were collected 
in a circular region close by with a
radius r=137$^{''}$. For bright afterglows with PC mode count rates $>$ 1 count
s$^{-1}$ the source photons were selected in an annulus that excludes the inner
pixels in order to avoid the effects of pileup.
For Windowed Timing mode \citep[WT; ][]{hillj04} data
we extracted
source and background photons in boxes with a length of 40 pixel each, except from very
bright bursts like e.g. GRB 060729 \citep{grupe07} for which we applied the method as
described in \citet{romano06}.
For spectral fitting of the PC and WT mode data only events with 
grades 0-12 and 0-2, respectively, were
included.
The X-ray spectra  
were re-binned by {\it grppha} 3.0.0 having 20 photons per bin and  
analyzed by {\it XSPEC} version 12.3.0 \citep{arnaud96}.
The auxiliary response files (arfs) were created by {\it
xrtmkarf} using arfs version 008.
We used 
the standard response matrix swxpc0to12\_20010101v008.rmf for the PC mode data 
and swxwt0to2\_20010101v008.rmf for the WT data.
Note: These are standard reduction techniques as used in the first XRT 
refined analysis GCN circular from \swift\ bursts.

\section{\label{results} Results}

Table\,\ref{grb_list} lists the redshift, Galactic absorption column
density $N_{\rm H, gal}$, 
$\Delta N_{\rm H} = N_{\rm H, fit} - N_{\rm H, gal}$,  the intrinsic column
density $N_{\rm H, intr}$ at the redshift of the burst,
X-ray energy 
spectral slope \bx, $\chi^2/\nu$ of the power law fit with the
absorption column density as a free parameter ($N_{\rm H, free}$)
and fixed to the Galactic
value as given by \citet{dic90}, detection flag for the \swift~UVOT,
and the reference for the spectroscopic
redshift measurement. In cases where the free-fit
absorption column density $N_{\rm H, fit}$ is within the errors consistent with
the Galactic value, we set $\Delta N_{\rm H}$=0.

\begin{figure*}
\epsscale{1.5}
\plottwo{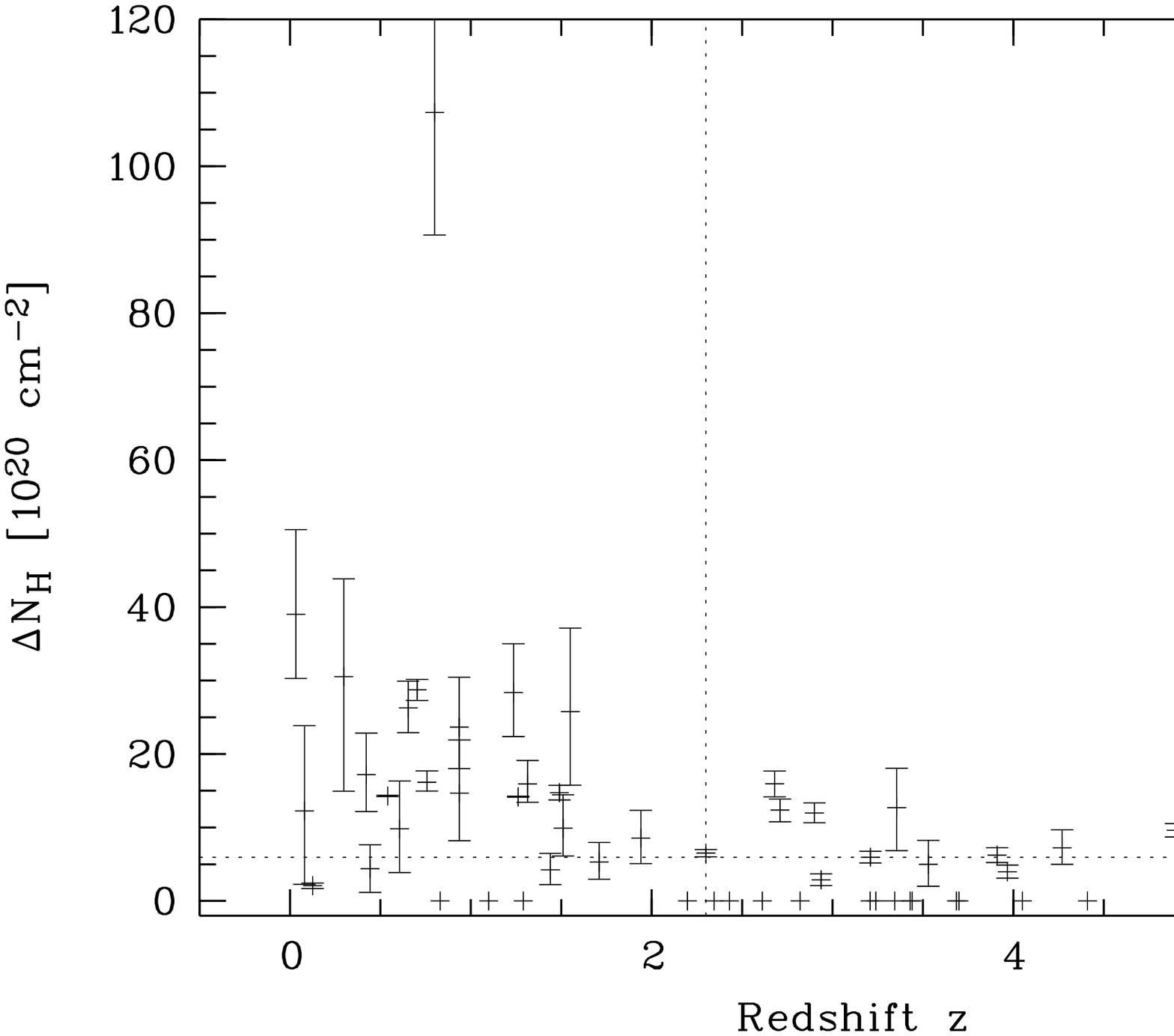}{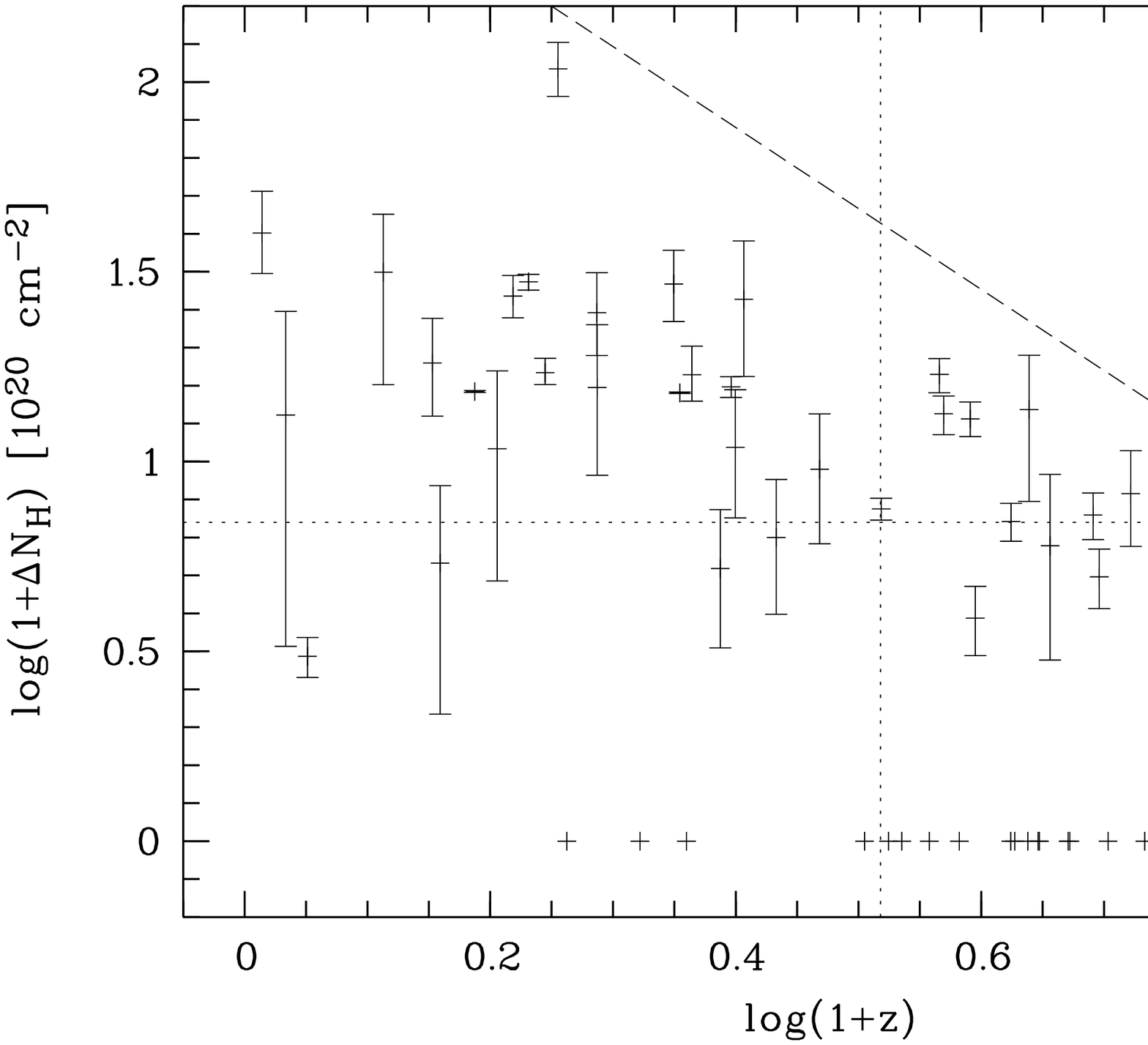}
\caption{\label{z_delta_nh} 
Redshift z vs. $\Delta N_{\rm H}$ relation among the \swift-observed bursts with
spectroscopic redshifts. The left panel displays the direct values while the right
panel shows the values with $log (1+z)$ and $log (1 + \Delta N_{\rm H})$.  
The
dotted lines display the median values for $\Delta N_{\rm H}=5.00\times 10^{20}$
cm$^{-2}$ and z=2.30 used for the 2$\times$2 contingency test (see text). The dashed
line in the right panel displays the line of the maximum expected value for the
redshift as given from equation 1.
}
\end{figure*}

The left panel of
Figure\,\ref{z_delta_nh} compares
redshift versus $\Delta N_{\rm H}$.
The dotted lines at z=2.30 and $\Delta N_{\rm H}=5.95\times 10^{20}$ cm$^{-2}$
are the medians in z and $\Delta N_{\rm H}$. These lines are used as 
cutoff lines to separate between low and high-redshift and low and high excess
absorption groups in a 
$2\times2$ contingency table\footnote{We evaluate the statistical
significance of our results by use of 2$\times$2 contingency
tables.
A 2$\times$2 contingency table is a statistical tool that compares a
property of two groups, such as in the present example: low- and
high-redshift bursts with or without significant additional absorption
above the Galactic column density. The way this method works is the following: 
assume a number of low redshift bursts  $n = l + k$, with $l$ of them having low and
$k$ having high column densities in excess of the Galactic value. The ratio of low to
high absorption column objects is then $l/k$. For the high-redshift bursts the
numbers are $z = x + y$ with $x$ representing
the number of bursts with low excess absorption
column densities and $y$ representing
 high excess absorption column densities. If the
ratio of low to high absorption bursts in high-redshift bursts is the same as
among low-redshift bursts, the number of high-redshift bursts  with high
absorption column densities is $y = k\times x/l$. 
The 2x2 contingency table compares the number of objects in each cell 
to the expected number under this assumption, and can be used to calculate 
the probability that the deviation
from the expected number of objects is just random.}.
The results of grouping these data by
these cutoff lines are given in Table\,\ref{2x2tab}. 
We can  immediately see  that bursts with
high excess absorption column densities $\Delta N_{\rm H}$ will most likely be at low
redshifts while afterglows with small $\Delta N_{\rm H}$ 
 are most likely at high redshifts.
From the $2\times2$ contingency table 
 a probability of only P=0.0011 (using a two-tailed Fisher
Exact Probability test) that this result is random can be calculated.

Another statistical test to check whether the relation is purely random 
 is the
Spearman rank order test.
 A Spearman rank order test
results in a correlation coefficient $r_s=-0.51$ with a Student's T-test
$T_s=-4.3$ and a probability P$< 10^{-4}$ of a random distribution.

The right panel of Figure\,\ref{z_delta_nh} displays the above results as $log (1+z)$
vs. $log(1+\Delta N_{\rm H})$. We use this diagram to conservatively
draw a line along the envelope of the  $log (1+z)$
vs. $log(1+\Delta N_{\rm H})$ distribution of the bursts, including the errors, in
order to determine maximum redshift of a burst with respect to its excess 
absorption $\Delta N_{\rm H}$. 
This line is displayed as the dashed line in the right panel of 
Figure\,\ref{z_delta_nh} and can be described as:

\begin{equation}
log(1+z) < 1.3 - 0.5\times log(1+\Delta N_{\rm H})
\end{equation} 

with $\Delta N_{\rm H}$ in units of $10^{20}$ cm$^{-2}$.
From this equation we can limit the maximum expected redshift based on the excess
absorption. Note that this equation does not estimate the redshift of a burst.
The equation only predicts the maximum redshift of a burst.
As examples, the redshift of 
a burst with an excess absorption $\Delta N_{\rm
H}=4\times 10^{21}$ cm$^{-2}$ is expected not to exceed z=2.1, $\Delta N_{\rm H}=2\times
10^{21}$ cm$^{-2}$ has z$<$3.4, and $\Delta N_{\rm H}=1\times 10^{21}$ 
has z$<$5.
Also note  that the number of bursts with redshifts
z$>$2.3 (the median redshift)
and no excess absorption detected, 
so the free-fit absorption column density $N_{\rm H.fit}$
is consistent with the Galactic value,
is three times as high as the number of bursts with no excess absorption and a
redshift z$<$2.3. In case we do not detect excess absorption in a burst,
most-likely this will be a burst with a redshift z$>$2.3.

Our findings of an observational relation between the excess absorption column
density $\Delta N_{\rm H}$ and redshift is due purely
to an observational artefact: we
only detect any excess absorption in high-redshift bursts which have a very large
intrinsic column density, as shown in Figure\,\ref{z_nh}. The dashed line at the
lower boundary of the $N_{\rm H, intr}$ - z distribution displays the maximum 
redshift at which an intrinsic column density can be detected in the XRT energy
window.

\section{\label{discuss} Discussion}

Our main result is that there is a clear
 anti-correlation between absorption column
density in excess of the Galactic value $\Delta N_{\rm H} = N_{H, fit} - N_{\rm
H, gal}$ and redshift.
This relation can be used to limit the range of possible redshifts. 
GRBs for which early
X-ray spectra are consistent 
with the Galactic absorption column density are most-likely at
higher redshifts ($z>2$), 
although a few examples of low-redshift GRBs also fall into
this category. However, GRBs with
 a significant excess column density of 
$\Delta N_{\rm H}>2 \times 10^{21}$ cm$^{-2}$  are
exclusively at redshifts z$<$2.0.

\begin{figure}
\epsscale{1.2}
\plotone{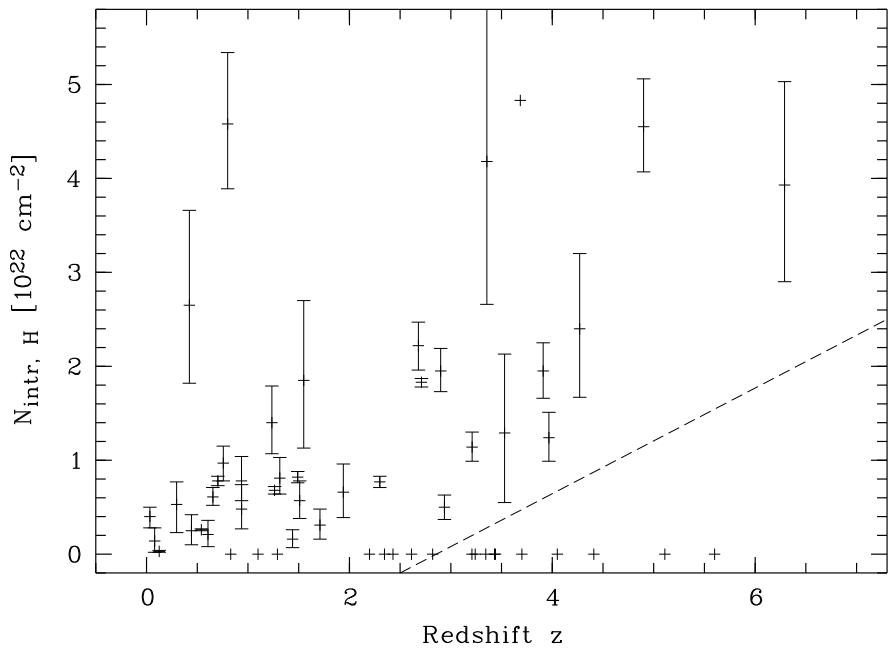}
\caption{\label{z_nh} Redshift z vs. 
the intrinsic column density $N_{\rm H, intr}$. The dashed line marks the
redshift limit at which an intrinsic column density can be detected in the XRT.
}
\end{figure}

The strong correlation we find between redshift and intrinsic 
$N_{\rm H}$
is an observational artefact. 
We can only detect an intrinsic absorber if the
absorption column density is large enough to affect the observable energy
range. For a z=4 burst the low energy cutoff of the detector at 0.3
keV is at 1.5 keV in the rest frame of the burst. In order to detect any
significant additional absorption in the observed energy window
above the Galactic value the
intrinsic (redshifted) absorption has to be in the order of at least
$10^{22}$ cm$^{-2}$, as shown in Figure\,\ref{z_nh}.

There are a few points one should be cautious about: 1) Our study is biased
towards afterglows which have spectroscopic redshifts and are therefore
detectable at optical and/or  NIR wavelengths. 
 As a result, we  2) may miss
afterglows in galaxies seen edge-on. These afterglows would suffer from significant
absorption columns on the order of several times $10^{23}$ cm$^{-2}$ as is
commonly observed in e.g.
Seyfert 2 galaxies. 3) Some of the early XRT WT mode data are not
well-fitted by a single absorbed power law model, e.g. GRBs 060614 and 060729
\citep[][ respectively]{mangano07, grupe07}. These GRBs display dramatic
changes in their X-ray spectra within minutes and the spectrum cannot be
modelled by one single power law. 
In some cases the spectra require multi-component spectral
models as it is the case for GRB 060729 \citep{grupe07}.

The immediate application of our column density - redshift relation is to
optimize ground-based optical and NIR follow up observations. We plan to include
the redshift limit information in future XRT refined analysis GCNs.
Photometric redshifts can also be calculated on the basis of UVOT data (Vanden Berk et al.,
in preparation). We tested our X-ray method with bursts that have UVOT photometric
redshifts, but no spectroscopic redshifts and found that the results agree with each other.
The advantage, however,  of
using the \swift-XRT data instead of the UVOT data to estimate a maximum redshift is
that 
\swift-XRT data are typically processed faster at the NASA \swift\ Data Center
than the UVOT data.
This is due to the larger data volume of the UVOT compared to the XRT data. 
Therefore a redshift prediction can be given faster on the basis of
the XRT data than on the UVOT data.  
This will
give ground-based observers at large telescopes a tool to decide which
spectrograph to use - an optical spectrograph for the low-redshift bursts and a
NIR spectrograph for the high-redshift bursts and will optimize
the use of large ground-based telescopes. Note, however, that the purpose of this paper
is not to discourage observers from obtaining spectra of bursts with  predicted low
redshifts. Each spectroscopic redshift - low or high redshift - is important, because it
enables us to determine physical parameters of the burst such as the isotropic energy
or the break times in the light curve. Therefore we encourage ground-based observers to
continue to obtain spectra of afterglows whenever possible. The larger the number of
bursts with spectroscopic redshifts the better our understanding of the physics of GRBs
will be. 
 
\acknowledgments

We would like to thank all observers at ground-based 
optical telescopes for their effort to obtain redshifts of the 
\swift\ afterglows.  We would also like to thank Sergio Campana 
for discussion related to the XRT calibration, Cheryl Hurkett 
for sending us a draft of her paper on GRB 050505, and Abe Falcone for various
discussions on the determination of the absorption column densities.
In particular we want to thank Eric Feigelson for various discussions on
statistics, and our referee Johan Fynbo for a fast and detailed referee's 
report that significantly improved the paper. 
This research has made use of data obtained through the High Energy 
Astrophysics Science Archive Research Center Online Service, provided by the 
NASA/Goddard Space Flight Center.
At Penn State we acknowledge support from the NASA Swift program
through contract NAS5-00136.



\clearpage

\clearpage

\begin{deluxetable}{llrrrccccl}
\tabletypesize{\tiny}
\tablecaption{List of all long burst GRB afterglows with \swift~observations and reported
redshifts \label{grb_list}}
\tablewidth{0pt}
\tablehead{
\colhead{GRB} & \colhead{z} & 
 \colhead{$N_{\rm H, gal}$\tablenotemark{2}} & 
 \colhead{$\Delta N_{\rm H}$\tablenotemark{3}} & 
 \colhead{$N_{\rm H, intr}$\tablenotemark{4}} & 
\colhead{\bx\tablenotemark{5}} &
 \colhead{$\chi^2/\nu$ \nh-free\tablenotemark{6}} &
 \colhead{$\chi^2/\nu$ \nh-fix\tablenotemark{7} }&
\colhead{UVOT}\tablenotemark{8} & 
\colhead{Redshift reference}  
} 
\startdata
050126  &  1.29    & 5.28 &  --- & --- &  1.01\plm0.18 & --- & 16/15  & n & \citet{berger05b} \\
050315  &  1.94    & 4.34 &  8.55$^{+3.80}_{-3.48}$ & 0.66$^{+0.30}_{-0.27}$ & 1.42$^{+0.18}_{-0.16}$ & 43/45 & 61/46 & n & \citet{berger05b} \\
050318  &  1.44    & 2.79 &  4.23$^{+2.24}_{-2.00}$ & 0.16$^{+0.10}_{-0.09}$ & 1.03$^{+0.11}_{-0.10}$ & 73/74 & 86/75 & y & \citet{berger05b} \\
050319  &  3.24    & 1.13 &  ---                    & ---                    & 1.10$^{+0.17}_{-0.15}$ & 27/36 & 31/37 & y &
\citet{jakobsson06}\tablenotemark{9} \\
050401  &  2.90    & 4.84 & 11.96$^{+1.39}_{-1.33}$ & 1.95$^{+0.22}_{-0.24}$ & 1.09$^{+0.06}_{-0.05}$ & 296/258 & 614/259  & n &
\citet{watson06}\tablenotemark{10} \\
050408  &  1.2357  & 1.74 & 28.36$^{+6.60}_{-5.97}$ & 1.40$^{+0.39}_{-0.33}$ & 1.31$^{+0.20}_{-0.19}$ & 39/36 & 120/37& y & \citet{berger05b} \\
050416A &  0.6535  & 2.07 & 26.28$^{+3.37}_{-3.64}$ & 0.61$^{+0.10}_{-0.09}$ & 1.20$^{+0.12}_{-0.11}$ & 65/84 & 320/85 & y &
\citet{jakobsson06c}\tablenotemark{11} \\
050505  &  4.27    & 2.04 &  7.23$^{+2.45}_{-2.25}$ & 2.40$^{+0.80}_{-0.73}$ & 1.18$^{+0.12}_{-0.11}$ & 51/77 & 85/78  & n &
\citet{jakobsson06c}\tablenotemark{12} \\
050525  &  0.606   & 9.10 &  9.81$^{+6.53}_{-5.96}$ & 0.21$^{+0.13}_{-0.15}$ & 1.08$^{+0.23}_{-0.21}$ & 23/27 & 31/28  & y &
\citet{jakobsson06c}\tablenotemark{13} \\
050603  &  2.821   & 1.19 &  ---                  & ---                      & 0.75\plm0.10           & 30/34 & 31/35  & y &
\citet{jakobsson06c}\tablenotemark{14} \\
050730  &  3.967   & 3.06 &  3.97$^{+0.91}_{-0.87}$ & 1.24$^{+0.25}_{-0.27}$ & 0.71\plm0.04           & 177/174 & 243/175 & y &
\citet{jakobsson06c}\tablenotemark{15} \\
050802  &  1.71    & 1.78 &  5.31$^{+2.66}_{-2.35}$ & 0.31$^{+0.17}_{-0.15}$ & 1.00$^{+0.12}_{-0.11}$ & 61/67 & 77/68  & y &
\citet{jakobsson06c}\tablenotemark{16} \\
050803  &  0.422   & 5.63 & 17.19$^{+5.65}_{-5.02}$ & 2.65$^{+1.01}_{-0.83}$ & 1.05$^{+0.19}_{-0.17}$ & 57/59 & 99/60  & n & \citet{bloom05} \\		 
050820  &  2.612   & 4.71 &  ---                    & ---                    & 0.00$^{+0.04}_{-0.04}$ & 87/100 & 87/101 & y &
\citet{jakobsson06c}\tablenotemark{17} \\
050824  &  0.83    & 3.62 &  ---                    & ---                    & 0.83$^{+0.24}_{-0.22}$ & 23/20 & 22/21  & y &
\citet{jakobsson06c}\tablenotemark{18} \\
050826  &  0.297   & 21.70& 30.54$^{+15.6}_{-13.3}$ & 0.53$^{+0.30}_{-0.24}$ & 0.93$^{+0.26}_{-0.23}$ & 29/25 & 45/26  & n & \citet{halpern06} \\
050904  &  6.29    & 4.93 &  4.05$^{+1.16}_{-1.12}$ & 3.93$^{+1.10}_{-1.03}$ & 0.47\plm0.04           & 126/87 & 104/92 & n & \citet{kawai05}\\
050908  &  3.344   & 2.14 &  ---                    & ---                    & 2.01$^{+0.41}_{-0.31}$ & 15/17 & 15/18 & y & \citet{jakobsson06c}\tablenotemark{19} \\
050922C &  2.198   & 3.43 &  ---                    & ---                    & 0.99$^{+0.11}_{-0.10}$ & 22/25 & 22/26 & y & 
\citet{piranomonte05}\tablenotemark{20} \\
051016B &  0.9364  & 3.64 & 23.67$^{+6.77}_{-5.65}$ & 0.78$^{+0.26}_{-0.21}$ & 1.13$^{+0.21}_{-0.19}$ & 30/32 & 93/33 & y & \citet{nardini06}\tablenotemark{21} \\
051022  &  0.80    & 4.06 &  107.3$^{+18.8}_{-16.7}$ & 4.58$^{+0.76}_{-0.69}$ & 1.38$^{+0.22}_{-0.20}$ & 35/47 & 206/48 & n & \citet{gal-yam05} \\
051109A &  2.346   & 17.5 &  ---                    & ---                    & 1.03\plm0.11           & 43/34 &  43/35 & y & \citet{nardini06}\tablenotemark{22} \\
051109B &  0.080   & 13.1 & 12.26$^{+11.6}_{-10.0}$ & 0.14$^{+0.14}_{-0.12}$ & 1.16$^{+0.33}_{-0.29}$ & 11/15 &  15/16 & n & \citet{perley05} \\
051111  &  1.55    & 5.02 & 25.77$^{+11.36}_{-10.01}$ & 1.85$^{+0.72}_{-0.85}$ & 1.77$^{+0.47}_{-0.41}$ & 16/12 &  36/13 & y & \citet{nardini06}\tablenotemark{23} \\
060115  &  3.53    & 12.60 & 5.00$^{+3.00}_{-3.25}$ & 1.29$^{+0.84}_{-0.74}$ & 0.92$^{+0.12}_{-0.11}$ & 78/85 & 86/86 & n & \citet{delia06} \\
060123  &  1.099   & 1.48 & ---                     & ---                    & 0.77$^{+0.26}_{-0.30}$ & 11/15 & 14/16 & n & \citet{berger06a} \\
060124  &  2.30    & 9.16 & 6.51\plm0.50            & 0.77\plm0.06           & 0.36\plm0.01           & 654/453 & 1215/454 & y & \citet{mirabal06a}\\
060206  &  4.05    & 0.94 & ---                     & ---                    & 1.25$^{+0.34}_{-0.29}$ & 9/11  & 13/12 & y & \citet{fynbo06} \\
060210  &  3.91    & 8.52 & 6.23$^{+1.03}_{-1.00}$  & 1.95$^{+0.29}_{-0.30}$ & 1.05\plm0.04           & 195/107 & 315/108 & n & \citet{cucchiara06a}\\
060218\tablenotemark{24}  
        &  0.033   & 11.00 & 39.01$^{+11.38}_{-8.73}$ & 0.40$^{+0.12}_{-0.10}$ & 3.13$^{+0.53}_{-0.38}$ & 17/26 & 90/27 & y & \citet{mirabal06b} \\
060223A &  4.41    & 5.93  & ---                    & ---                    & 0.84$^{+0.15}_{-0.13}$ & 15/14 & 16/15 & n & \citet{berger06b} \\
060418  &  1.49    & 9.27  & 14.27$^{+1.00}_{-0.99}$ & 0.82\plm0.06          & 1.22\plm0.04           & 218/174 & 1073/149 & y &
\citet{ellison06}\tablenotemark{25} \\
060502A &  1.51    & 2.97  &  9.90$^{+4.56}_{-3.79}$ & 0.57$^{+0.21}_{-0.19}$ & 2.36$^{+0.46}_{-0.38}$ & 64/35 & 89/36 & y & \citet{cucchiara06b} \\
060510B &  4.90    & 3.78  &  9.61$^{+0.92}_{-0.89}$ & 4.55$^{+0.51}_{-0.48}$ & 0.38\plm0.03          & 194/183 & 238/183 & n & \citet{price06} \\
060512  &  0.4428  & 1.43  &  4.40\plm3.24           & 0.25\plm0.17           & 3.07\plm0.38          & 22/10   & 23/11   & y & \citet{bloom06} \\
060522  &  5.11    & 4.83  &  ---                    & ---                    & 0.70$^{+0.19}_{-0.18}$ & 17/17 & 17/18 & y & \citet{cenko06} \\	
060526  &  3.21    & 5.46  &  5.95$^{+0.81}_{-0.78}$ & 1.14$^{+0.16}_{-0.15}$ & 0.79\plm0.03      & 217/158  &  408/159  & y & 
  \citet{berger06c}\tablenotemark{20} \\
060604  &  2.68    & 4.55  & 15.96$^{+1.72}_{-1.79}$ & 2.22$^{+0.25}_{-0.26}$ & 1.39\plm0.07          &  83/51 & 392/52 & y & \citet{castro06} \\
060605  &  3.711\tablenotemark{26}     & 5.11  &  ---                    & ---                    & 1.03\plm0.09           &  31/35 & 31/36  & y 
& \citet{still06}\tablenotemark{27} \\
060607  &  2.937   & 2.67  &  2.87$^{+0.82}_{-0.79}$ & 0.50\plm0.13           & 0.79\plm0.04           &  200/188 & 240/189 & y & \citet{ledoux06} \\
060614  &  0.125   & 3.07  &  2.07\plm0.37           & 0.03\plm0.01           & 0.45\plm0.15           &  446/256 & 541/257 & y &
\citet{mangano07}\tablenotemark{28} \\
060707  &  3.43    & 1.76  &  ---                    & ---                    & 0.94\plm0.12           & 13/18    & 13/9    & y & \citet{jakobsson06} \\
060714  &  2.71    & 6.72  & 12.36$^{+1.59}_{-1.52}$ & 1.83$^{+0.05}_{-0.04}$ & 0.98$^{+0.06}_{-0.05}$ & 261/252  & 501/253 & y & \citet{jakobsson06} \\
060729  &  0.54    & 4.82  & 14.29$^{+0.09}_{-0.08}$ & 0.26\plm0.01           & 1.76\plm0.04           & 505/297 & 1817/298 & y &
\citet{grupe07}\tablenotemark{29} \\
060904B &  0.703   & 12.10 & 28.80$^{+1.45}_{-1.40}$ & 0.78\plm0.05           & 1.16\plm0.04           & 540/425 & 2570/426 & y & \citet{fugazza06} \\
060906  &  3.685   &  9.66 & ---                     & ---                    & 1.51$^{+0.62}_{-0.51}$ &  12/10  &   16/11  & n & \citet{jakobsson06} \\
060908  &  2.43    &  2.73 & ---                     & ---                    & 1.30$^{+0.28}_{-0.25}$ &  29/25  &   31/26  & y & \citet{rol06} \\
060912  &  0.937   &  4.23 & 14.67$^{+7.25}_{-6.47}$ & 0.48$^{+0.26}_{-0.21}$ & 1.17$^{+0.25}_{-0.23}$ &  11/20  &   27/21  & y & \citet{jakobsson06} \\
060926  &  3.208   &  7.30 & ---                     & ---                    & 0.97$^{+0.33}_{-0.30}$ &  11/15  &   16/16  & y & 
    \citet{piranomonte06}\tablenotemark{30} \\
060927  &  5.6     &  5.20 & ---                     & ---                    & 0.82$^{+0.30}_{-0.27}$ &  11/15  &   11/16  & n & \citet{fynbo06b} \\
061007  &  1.261   &  2.22 & 11.87\plm0.04           & 0.52\plm0.02           & 1.01\plm0.01           & 1072/587 & 4784/588 & y & 
    \citet{schady07}\tablenotemark{31} \\
061110A &  0.757   &  4.87 & 16.15$^{+1.20}_{-1.56}$ & 0.97$^{+0.19}_{-0.18}$ & 2.05$^{+0.11}_{-0.10}$ & 242/197  &  579/198 & n & \citet{fynbo06c} \\
061110B &  3.44    &  4.83 & ---                     & ---                    & 1.04$^{+0.41}_{-0.37}$ &  4/3     &    4/4   & n & \citet{thoene06b} \\
061121  &  1.314   &  5.09 & 15.93$^{+2.50}_{-3.20}$ & 0.81$^{+0.22}_{-0.17}$ & 0.25\plm0.07           & 290/286  &  339/287 & y &
\citet{page07}\tablenotemark{32} \\
061222B &  3.355   & 27.70 & 12.70$^{+5.86}_{-5.35}$ & 4.18$^{+1.81}_{-1.53}$ & 1.60$^{+0.20}_{-0.18}$ & 78/47    &   95/47  & n & \citet{berger06d} 
\enddata

\tablenotetext{1}{Times after the burst that were selected from the event file}
\tablenotetext{2}{Absorption column densities are given in units of 10$^{20}$ 
cm$^{-2}$, Galactic \nh~values taken from \citet{dic90}}
\tablenotetext{3}{$\Delta N_{\rm H} = N_{\rm H, fit} - N_{\rm H, gal}$
in units of 10$^{20}$  cm$^{-2}$. A dash indicates that the $N_{\rm H}$ is
consistent with the Galactic value.}
\tablenotetext{4}{Column densities of the intrinsic absorber at the redshift of
the burst given in units of 10$^{22}$  cm$^{-2}$.}
\tablenotetext{5}{X-ray spectral slopes determined from an absorbed power law fit with the
absorption parameter left free, except for those sources where the a free absorption was below the
Galactic value. In those cases we fixed the absorption parameter to the Galactic value.}
\tablenotetext{6}{$\chi^2/\nu$ of a power law fit with the absorption column density as a free parameter}
\tablenotetext{7}{$\chi^2/\nu$ of a power law fit with the absorption column density fixed to the Galactic value}
\tablenotetext{8}{UVOT detection in any of the 7 filters including white light with y=detection and n=no detection}
\tablenotetext{9}{Redshift originally reported by \citet{fynbo05a}.}
\tablenotetext{10}{Redshift originally reported by \citet{fynbo05b}.}
\tablenotetext{11}{Redshift originally reported by \citet{cenko05}.}
\tablenotetext{12}{Redshift originally reported by \citet{berger05b}.}
\tablenotetext{13}{Redshift originally reported by \citet{foley05}.}
\tablenotetext{14}{Redshift originally reported by \citet{berger05c}; 
see also \citet{grupe06}}
\tablenotetext{15}{Redshift originally reported by \citet{chen05}.}
\tablenotetext{16}{Redshift originally reported by \citet{fynbo05c}.}
\tablenotetext{17}{Redshift originally reported by \citet{prochaska05}.}
\tablenotetext{18}{Redshift originally reported by \citet{fynbo05b}.}
\tablenotetext{19}{Redshift originally reported by \citet{fugazza05}.}
\tablenotetext{20}{See also \citet{jakobsson06}}
\tablenotetext{21}{Redshift originally reported by \citet{soderberg05}.}
\tablenotetext{22}{Redshift originally reported by \citet{quimby05}.}
\tablenotetext{23}{Redshift originally reported by \citet{hill05}.}
\tablenotetext{24}{GRB 060218/SN 2006aj has a very unusual initial spectrum that can not be fitted by a single power law.
It requires a strong thermal component \citep{campana06b}.
Therefore we only used the pc mode data for this unusual burst.}
\tablenotetext{25}{Redshift originally reported by \citet{dupree06}.}
\tablenotetext{26}{This redshift was revised by \citet{savaglio07} to z=3.78.}
\tablenotetext{27}{See also \citet{ferrero06}.}
\tablenotetext{28}{Redshift originally reported by \citet{price06b}.}
\tablenotetext{29}{Redshift originally reported by \citet{thoene06}.}
\tablenotetext{30}{Improved redshift given by \citet{delia06b}}
\tablenotetext{31}{Redshift originally reported by \citet{osip06} but also 
see \citet{jakobsson06b}.}
\tablenotetext{32}{Redshift originally reported by \citet{bloom06b}.}
\end{deluxetable}

\begin{deluxetable}{lrr}
\tablecaption{2$\times$2 contingency table results for the cuts at a redshift
z=2.30 and $\Delta N_{\rm H}=5.95\times 10^{20}$ cm$^{-2}$. 
\label{2x2tab}}
\tablewidth{0pt}
\tablehead{
 & \colhead{z$\leq$2.30} & 
\colhead{z$>$2.30}  
} 
\startdata
$\Delta N_{\rm H}\geq 5.95 \times 10^{21}$ cm$^{-2}$   & 20 &  7  \\
$\Delta N_{\rm H}< 5.95 \times 10^{21}$ cm$^{-2}$      &  8 & 20 
\enddata

\end{deluxetable}

\end{document}